\documentclass[onecolumn,showpacs,preprintnumbers,amsmath,amssymb]{revtex4}
\usepackage{graphicx}
\usepackage{subfigure}
\begin{document}
\title{Fluorescence intermittency in blinking quantum dots: renewal
or slow modulation? }
\author{Simone Bianco}
\email{sbianco@unt.edu}
\author{Paolo Grigolini}
\email{grigo@unt.edu}
\author{Paolo Paradisi}
\email{paradisi@le.isac.cnr.it}
\affiliation{Center for Nonlinear Science, University of North Texas,
P.O. Box 311427, Denton, Texas 76203-1427, USA}
\affiliation{Istituto dei Processi Chimico Fisici del CNR, Area della
Ricerca di Pisa, Via G. Moruzzi, 56124, Pisa, Italy}
\affiliation{Dipartimento di Fisica "E.Fermi" - Universit\`a di Pisa,
Largo Pontecorvo, 3 56127 PISA}
\affiliation{ISAC-CNR, sezione di Lecce, strada provinciale
Lecce-Monteroni, km 1.2, 73100, Lecce}
\date{\today}
\begin{abstract}
We study time series produced by the blinking quantum dots, by means
of an aging experiment, and we examine the results of this experiment
in the light of
two distinct approaches to complexity, renewal and slow modulation.
We find that the renewal approach
fits the result of the aging experiment, while the slow modulation
perspective does not. We make also an attempt at establishing the
existence of an intermediate condition.
\end{abstract}


\maketitle

\section{introduction}\label{introduction}
In the last few years there has been an increasing interest for the
intermittent fluorescence of new nanomaterials
\cite{experimental1,experimental2}, which for this reason have been
called \cite{blinkingquantumdots} \emph{Blinking Quantum Dots} (BQD).
The theoretical interest for these new materials is due, to a great
extent, to the properties of  their power spectrum. Pelton, Grier and
Guyot-Sionnest \cite{oneoverf}, as well as Chung and Bawendi \cite
{collection}, have studied macroscopic samples of quantum dots, and
have proved that they generate $1/f$ noise. Thus, understanding these
new materials is a problem directly connected with the origin of the
$1/f$ noise, an issue that according to the proponents of Self
Organized Criticality (SOC) should be settled using their theoretical
perspective \cite{bak}. The search for the dynamical origin of
complexity, meant to be a statistical condition departing from the
Poisson prescription, is a hot topic, and recently another proposal
of approach to complexity,
called \emph{superstatistics}, was made by Beck and Cohen \cite{cohenbeck}
(see also Ref.\cite{cohen}). On the other hand, there is an
increasing conviction
\cite{subordination1,subordination2,subordination3,subordination4}
that anomalous diffusion, and especially sub-diffusion, might find a
satisfactory theoretical framework in the \emph{subordination}
perspective.  In this paper we shall illustrate the subordination
perspective with physical arguments, and shall refer to this
theoretical approach simply as subordination. Here we limit ourselves
to saying that according to the original work of Montroll and Weiss
\cite{pioneeringwork}, the time distance between two consecutive jump
events, generating the walker diffusion,  is not fixed but is
randomly drawn from a distribution of waiting times.  This is a case
used to generate a new form of diffusion (anomalous), through
subordination to the ordinary diffusion. The
subordination perspective is also be applied to generate anomalous
fluctuation-dissipation processes. In this case the original events
are processes where diffusion is balanced by friction. By assuming
that the time distance between two consecutive events of this kind is
not fixed, we derive a non-ordinary fluctuation-dissipation  process
subordinated to the ordinary one \cite{failla}.

Subordination
yielding anomalous transport can be imagined as a theoretical
perspective where the collisional events generating randomness are
rare and separated by large time intervals with a non-exponential
distribution. This important property requires, in turn, a
physical explanation that hopefully might relate the deviation from
exponential distribution to the cooperation among the elementary
constituents of the system. The subordination perspective is based on 
the assumption that the phenomenon under study  is a \emph{renewal} 
one, insofar the jump occurrence is assumed to reset
to zero the system's memory.

In the case of BQD fluorescence,
many researchers have been studying the microscopic origin of
intermittency, and, on top of that, of why the intermittency of these
materials is characterized by non-Poisson distribution of times of
sojourn in the $``on"$, with the system emitting light, and 
\emph{``off"} state, with the system emitting no light
\cite{nesbitt,nesbittmodulation,hiddenmodulation,shinya}. To first
sight, all these papers seem to share the same theoretical
perspective as superstatistics. According to Cohen \cite{cohen}, the
process yielding the deviation from the canonical distribution is
realized by the system as follows. The particle resides in a cell in local
equilibrium  for a time extended enough as to reach the corresponding
canonical equilibrium, then it moves to another condition of local
equilibrium, and so on, in such a way that the resulting distribution
will turn out to be distinctly non-exponential, insofar as the
superposition of many exponential functions with different probabilistic
weights is not an exponential function. The sojourn in a condition of local
equilibrium must be extended in time enough for the system to adapt
itself to this condition. For this reason superstatistics is
equivalent to a form of very slow modulation.

The model discussed in Ref. \cite{nesbitt}  might be thought of as a form of
superstatistics. In fact, these authors propose a two-state model
with barrier height or barrier width fluctuation. This automatically
yields a waiting time distribution as a sum of infinitely many
exponential functions, one exponential for each of the barrier
heights or widths. However, as the authors point out, these
fluctuations cannot be slow. In fact, the case of slow fluctuations
would produce a correlation between two successive on-times or two
successive off-times, while the analysis of Ref. \cite{nesbitt} shows
that no significant correlation of this kind exists. Similar remarks
apply to the model of Refs.
\cite{nesbittmodulation,hiddenmodulation,shinya}. In all these models
there are traps with exponential waiting times distribution
densities. However, the system resides in each exponential trap
corresponding to a given state, for instance the state ``off", only
once, and it jumps back to the state ``on" before being trapped again
in the state ``off". This property ensures the renewal property. In
conclusion, superstatistics is a form of very slow modulation,
whereas renewal is compatible with a theory based on non-exponential
distribution densities emerging from the superposition of many
exponential waiting time distributions, provided that the system
\emph{does not} adopt the same exponential well for too many
successive trapping events.

As an example of physical process that
might be conveniently described by the slow modulation perspective,
we quote the interesting recent papers \cite{klafter1} and
\cite{klafter2}. These authors find \cite{klafter1} that their
experimental result cannot be described by the two-state semi-Markov
models proposed in Refs. \cite{verberknotworking,MarBar04}, being
therefore incompatible with the renewal perspective. The
chronological ordering of the off waiting times suggests the
occurrence of
a modulation corresponding to molecular conformational
changes.

It is therefore convenient to develop a technique of
analysis of the experimental data that might help the investigators
to establish the real nature of the process under study, namely whether
renewal, slow modulation, or an intermediate condition, applies. This
is the main purpose of this paper.
We shall show that, although
renewal and superstatistics can be used to produce the same waiting
time distribution density, the former approach to complexity
generates renewal aging, while the latter does not. The aging technique
has already been adopted by Brokmann \emph{et al.} \cite{brokman} to
prove that the physics of BQD is characterized by renewal. In this
paper, in addition to confirming the conclusions of this earlier
work, using the same aging technique, we make also an attempt at
assessing if a condition intermediate between renewal and very slow
modulation (superstatistics) might exist.

The outline of the paper
is as follows. In section \ref{renewal0} we show that renewal rests
on resetting the system's memory after any event (collision). In
Section \ref{modulation} we define modulation as a condition of time dependent
rate, with no renewal. We explain the form of modulation adopted in
this paper, and we show that
in the fast condition it becomes
compatible with non-Poisson renewal. We devote Section \ref{aging} to
reviewing the phenomenon of renewal aging and we explain why the
condition of very slow modulation should yield no aging. In Section
\ref{artificialdata}, with the help of
artificial sequences, we
illustrate the aging experiment that we propose for the analysis of
experimental sequences. At the same time we explore the unknown region
between renewal and very slow modulation. We examine and discuss
real experimental data in Section \ref{realexperiment}. Finally, we
devote Section \ref{final} to concluding remarks.

\section{renewal}\label{renewal0}

Let us consider a two-state renewal process, and, for simplicity,
let us assume that the distribution of sojourn times  in the
state ``on" is the same as the distribution in the state ``off".
This
assumption will prevent us from discussing with our techniques the
interesting effect recently discussed by
Verberk \emph{et al.}
\cite{interestingeffect}. These authors discussed the case where both
distributions have an inverse power law form with different indexes,
$\mu_{off}$ and $\mu_{on}$ and found that a Gibbs ensemble of
trajectories moving from the beginning of the state "on", produce a
fluorescence intensity decaying in time as a function proportional to
$1/t^{\mu_{off}-\mu_{on}}$, with $\mu_{off} > \mu_{on}$. We leave the
discussion of this interesting effect as a subject for future work.

We  assign to the Survival Probability (SP) of this process,
$\Psi(t)$, the inverse power law form
\begin{equation}
\label{experimental} \Psi(t) = \left(\frac{T}{T +
t}\right)^{\mu-1},
\end{equation}
with $\mu > 1$. This corresponds to the joint action of the time
dependent rate \cite{cox} $r(t) = r_{0}/(1 + r_{1} t)$, with
$r_{0} = (\mu-1)/T $ and $r_{1} = 1/T$, and of a resetting
prescription. To illustrate this condition, let us imagine the random
drawing of a number from
the interval $I = [0,1]$ at discrete times $i= 0,1,2...$. The
interval $I$ is divided into two parts, $I_{1}$ and $I_{2}$, with
$I_{1}$ ranging from $0$ to $p_{i}$, and  $I_{2}$ from $p_{i}$ to
$1$. Note that $p_{i} = 1 - q_{i} < 1$ and $q_{i}<<1$, and, as a
consequence, the number of times we keep drawing numbers from
$I_{1}$, without moving to $I_{2}$, is very large.

Let us evaluate the distribution of these persistence times, and
let us discuss under which conditions we get the SP of Eq.
(\ref{experimental}). The SP function is the probability of
remaining  in $I_{1}$ after $n$ drawings, and is consequently
given by
\begin{equation}
\label{towardscox0} \Psi(n) = \prod_{i=1}^{n} p_{i}.
\end{equation}
Using the condition $q_{i} << 1$, and evaluating the logarithm of
both terms of Eq. (\ref{towardscox0}), we obtain:
\begin{equation}
\label{towardscox} log (\Psi(n)) = - \sum_{i=1}^{n}  q_{i}.
\end{equation}
The condition $q_{i} << 1$
implies that $i$ and $n$ of Eq. (\ref{towardscox}) are so large as to
make $q_{i}$ virtually identical to a function of the continuous
time $t$,  $q_{i}  = r(t) = r_{0} \eta(t) $, with $\eta(t) = 1/(1 +
r_{1} t)$. Therefore $r(t)$ is a time dependent rate, with resetting as
we hereby shall see.
Thus, Eq. (\ref{towardscox}) yields the SP of Eq.
(\ref{experimental}), and the corresponding waiting
time distribution density, $\psi(\tau)$, reads
\begin{equation}
\label{theoretical} \psi(\tau) = (\mu - 1) \frac{T^{\mu -1}}{(\tau +
T)^{\mu}}.
\end{equation}

We denote as \emph{collisions} the rare drawings of a number from
$I_{2}$, followed by resetting. Thus the collisions occurring at
times $\tau_{1}$, $\tau_{1} + \tau_{2}$, ..., yield: $\eta(t) = 1/(1
+ r_{1} t), 0<t<\tau_{1}$; $\eta(t) = 1/(1 + r_{1} (t-\tau_{1})),
\tau_{1} <t<\tau_{1} + \tau_{2}$, and so on. Note that $\eta(0) = 1$
means that we \emph{prepare} the system at time $t = 0$.  We might
adopt a coin tossing
prescription to decide whether to keep or to change sign, after any
collision. However, in this paper, as earlier pointed out, we
do not pay attention to the problem of fluorescent intensity changing
in time as an effect of ensemble average.  Thus, for
simplicity, our theoretical remarks  refer  to a sequence
$\{\tau_{i}\}$, where
the times $\tau_{i}$ are randomly drawn so as to yield the analytical
form of Eq. (\ref{theoretical}), without assigning to them either an
``on" or an ``off" symbol.

\section{modulation}\label{modulation}
The renewal condition
described in Section \ref{renewal0} must not be confused with the
case of a time dependent rate, with no renewal. The time dependence
of $r(t)$, with no renewal, might obey a deterministic or a
stochastic prescription. An example of the former case is
$q(t) = A
+ B cos (\omega t)$, with no renewal. We think that the physical
process studied by the authors of Refs. \cite{klafter1,klafter2},
might be adequately described by a prescripion of this kind, not
necessarily periodic, or quasi periodic, reflecting however the
molecular conformational changes in time.

The specific cases
discussed in this paper are closer in spirit to the condition of
stochastic dependence on time.
This means that $r(t)$ is a stochastic
function of time, so that we have to study:

\begin{equation}
\label{fluctuatingcase}
\Psi(t) = <exp \left(-\int_{0}^{t}dt' r(t')\right)>.
\end{equation}

An interesting example of treatment of this kind is offered by the
recent work of Brown \cite{brown}. It has to be pointed out that the
evaluation of the characteristic function of
Eq.~\eqref{fluctuatingcase}  might be a difficult problem, but in the
limiting cases of very fast or very slow modulations. The first
condition departs from the non-Markov condition of interest for us in
this paper. The latter condition does not, and can be adopted to
derive the non-exponential behavior of $\Psi(t)$.

Let us assume that the fluctuation $r(t)$ has an equilibrium
distribution, $p_{eq}(r)$. In the special case where the time scale
of the fluctuation $r(t)$ is virtually infinite, the SP $\Psi(t)$ becomes:
\begin{equation}
\label{infinitelyslowfluctuations1}
\Psi(t) = \int_{0}^{\infty} dr p(r) exp(-r t).
\end{equation}
In fact, let us consider a Gibbs ensemble of identical systems,
obeying Eq.~\eqref{fluctuatingcase}. At the moment when the
observation begins, at time $t=0$, each of these systems has a rate
$r$, given by the distribution $p_{eq}(r)$. If the time scale of the
rate fluctuations is virtually infinite, during the observation
process each system will keep unchanged its own rate, thereby
producing Eq.~\eqref{infinitelyslowfluctuations1}. This makes this
picture essentially identical to the approach to complexity recently
proposed by Beck \cite{beck}.

According to a mathematical prescription borrowed from Beck
\cite{beck}, we find that the analytical expression of
Eq.~\eqref{experimental} is recovered by using the following form:
\begin{equation}
\label{christian}
p_{eq}(r)  =  \frac{T^{\mu-1}}{\Gamma(\mu-1)} r^{\mu-2} exp(-r T).
\end{equation}
This proposal was used in a later paper \cite{cohenbeck} to develop a
new approach to complexity, denoted as superstatistics.
This approach  is attracting the attention of an increasing number of
researchers (see, for instance, \cite{stupid1,stupid2,stupid3}), and
for this reason is worth of consideration.

Note that once the
inverse power law form of Eq. (\ref{experimental}) for the SP
$\Psi(t)$ is realized, the corresponding waiting time
distribution density is given by
\begin{equation}
\label{infinitelyslowfluctuation}
\psi(t) = \int_{0}^{\infty} dr p(r) r exp(-r t).
\end{equation}

In principle, the adoption of the time-dependent rate
prescription of Section \ref{renewal0} would make it possible to
describe modulation processes of any kind. It would be enough to set
$r_{1} = 0$ and replace $r_{0}$ with $q(t)$, and make it either
deterministic or stochastic. However, to generate the time series
$\{\tau_{i}\}$ corresponding to the modulation prescription, fast,
slow or of intermediate speed, we adopt a different procedure. We select from
the distribution of Eq. (\ref{christian})
the sequence $\{r_{j}\}$. For any exponential waiting time
distribution $\psi_{j}(t) = r_{j} exp(-r_{j} t)$, we select
$N_{m}^{j}$ times,  $\tau^{j}_{1}, ...., \tau^{j}_{N_{m}^{j}}$.
A more
realistic picture might allow $N_{m}^{j}$ to fluctuate. We expect,
however, that for ordinary fluctuations about a common
value
$<N_{m}^{j}> = N_{m}$, the physics of the process does not change.
Thus, for simplicity we assign to all the numbers $N_{m}^{j}$ the same
value $N_{m}$. The time series to compare to the one derived
according to the prescription of Section \ref{renewal0} is defined by
$\{\tau_{i}\} = \tau_{1}^{1},....\tau_{N_{m}}^{1},
\tau_{1}^{2},....\tau_{N_{m}}^{2}, ...$. The benefit of this
criterion is that $N_{m} = 1$ makes the resulting time series
equivalent to that generated by the models of
Refs.\cite{nesbitt,nesbittmodulation,hiddenmodulation,shinya} and,
consequently, to the renewal prescription of Section \ref{renewal0}. $\ \ $ \\

It is important to stress that the ideal condition of totally
renewal process and the ideal condition of infinitely slow modulation
are characterized by a marked
difference concerning ergodicity.  This important issue has been
recently discussed by Margolin and Barkai \cite{margolin}. In the
totally renewal case, with $\mu < 2$, the system does not admit any
stationary condition \cite{ignaccolo} and the stationary correlation
function does not exist. The non-stationary correlation function can
be defined making an ensemble average, as illustrated, for instance,
by the authors of Ref. \cite{lastlast}.
The authors of Ref.
\cite{margolin}, on the contrary, adopt a single trajectory picture
and study the time averaged intensity correlation function of the BQD
signal, supposed to be totally renewal,  as well as with identical ``on"
and ``off" distribution, as assumed in this paper. The correlation
function is a stochastic property characterized by U- and W-shaped
distributions.

The case of infinitely slow modulation would suggest
the adoption of an ensemble rather than individual trajectory
treatment.  However, in this paper
we adopt the individual
trajectory treatment also for the case of slow, but not infinitely
slow, modulation. It is expected that in this case ergodicity is not
violated, in a striking contrast with the condition of total renewal
\cite{margolin}.

We think that moving from $N_{m} = 1$, where the
properties found by Margolin and Barkai \cite{margolin} apply,  to
$N_{m} = \infty$, where only the ensemble treatment is possible,
implies the exploration of an unknown region, of which this paper
affords a preliminary treatment.
For the reader to appreciate this
aspect, we would like
to introduce the concept of pseudo event. This
concept is similar to that
proposed in an earlier publication \cite{memorybeyondmemory}. The
authors of this paper \cite{memorybeyondmemory} found that in some
problems of medical interest
the connection between scaling and waiting time distribution does not
correspond to the prescription of the renewal theory. This is so as a
consequence of the fact that  the times of the series under study
turned out to be correlated~\cite{memorybeyondmemory}.

In the case
of modulation, we define as pseudo events all the drawings of  waiting
times from the same Poisson distribution, after the first drawing. In
the case $N_{m} = 1$
there are no pseudo events. In the case $N_{m} = 2$ there is one
pseudo event, and so on. The quantity $N_{m}-1$ defines the number of
pseudo events per critical event. By critical event, we denote the
drawing of a given Poisson parameter $r$. In practice, the occurrence
of a critical event corresponds to the first drawing of a waiting
time from a Poisson distribution with rate $r_{j}$, namely the time
$\tau^{j}_{1}$.

The drawing of the next waiting times from
the same distribution, implies a subtle deviation from renewal. This
form of correlation is not easy to detect. In fact, although
consecutive sojourn times are drawn from the same Poisson
distribution, they are by definition independent  one from the
other. If the time correlation function is $<\tilde\tau_{i}
\tilde\tau_{j}>$, with $\tilde \tau \equiv \tau - <\tau>$ for a
finite portion of the sequence, we
expect it to yield $(<\tau^{2}> -
<\tau>^{2}) \delta_{i,j}$, with
$\delta_{i,j}$ is the delta of
Kronecker.

It is a striking and surprising fact that the
correlation produced by modulation is invisible to the ordinary
correlation  test. This is so because the sojourn times,
although derived for an extended period of time from the same Poisson
distribution, are randomly drawn. The aging experiment reveals this
subtle form of correlation. We refer to the times that are correlated
the ones to the other as pseudo events, regardless the origin of this
correlation, which might occur in the form discussed in the earlier
publication \cite{memorybeyondmemory} or in the even more subtle form
of this paper. We think that the aging reduction
depends on the ratio of  pseudo to critical events, regardless the origin
of correlation. In this paper we find that the aging reduction
depends indeed on the ratio of pseudo to critical events, of the type
here introduced.  On the basis of this result with artificial
sequence, we make also
an attempt at evaluating the amount of pseudo events that might be
present in the real BQD time series.

\section{Aging and modulation }\label{aging}
For an intuitive
description of the concept  of aging we shortly review the treatment
of the earlier work of Ref. \cite{earlier0}. The renewal condition of
Section \ref{renewal0} can also be realized with the following
dynamical model. A particle moves in the interval $I \equiv [0,1]$
driven by the equation of
motion
\begin{equation}
\label{52}
\frac{dy}{dt} = \alpha
y^{z},
\end{equation}
with $0 < \alpha << 1$ and $z>1$.
When it
reaches the point $y = 1$ it is injected back in a random position between $0$
and $1$ with uniform
probability, thereby producing another extended time of sojourn
within the interval $I$. The connection between the waiting time
distribution density and
the uniform initial distribution is given
by
\begin{equation}
\label{fundamental}
\psi(t)dt =
p(y_{0})dy_{0}.
\end{equation}
It is easy to prove that the resulting
time distribution density is given again by Eq. (\ref{theoretical})
with
\begin{equation}
\mu \equiv \frac{z}{z
-1}
\end{equation}
and
\begin{equation}
T  \equiv \frac{\mu -1
}{\alpha}.
\end{equation}

The uniform back injection is equivalent
to the resetting prescription of Section \ref{renewal0}, and, in
fact, this model is renewal, and it is equivalent to the model of
Section \ref{renewal0}, but its adoption in this section serves
better the purpose of explaining renewal aging and the lack of aging
in the case of very slow modulation.

The waiting time distribution
density given by Eq. (\ref{fundamental}) corresponds to beginning the
observation at the moment when the system is prepared in the uniform
distribution $p(y_{0}) = 1$. As a result of the injection back
process this distribution changes upon time change. If the
observation of the first times of sojourn is made at a later time
$t_{a} > 0$, the corresponding waiting time distribution density is
given
by
\begin{equation}
\psi_{t_{a}} (t)dt = p(y_{0},
t_{a})dy_{0}.
\end{equation}
The dependence of $\psi_{t_{a}}(t)$ on
$t_{a}$ is the renewal aging that we want to assess in this paper by
means of a suitable numerical experiment.
An exact expression for
$\psi_{t_{a}}(t)$ is available \cite{godreche,earlier0}, but, since it is not
expressed as a simple analytical formula, it is not suitable for the
practical purposes of this paper. For
this reason, we prefer to adopt the expression:
\begin{equation}
\label{approximated}
\psi_{t_{a}}(t) =  \frac{\int_{0}^{t_{a}} dy \psi (t + y)}{K_{t_{a}}},
\end{equation}
where $K_{t_{a}}$ is a suitable normalization constant. The meaning
of this approximated expression is evident. We assume that the first
sojourn times observed might have begun prior to $t = t_{a}$,
anywhere between $t = 0$ and $t=t_{a}$, with the restrictive
condition that the earlier laminar region, only one, began at $t =
0$. Actually, the last laminar region might be at the end of a sequel
of an arbitrarily large number of jumps, thereby generating
corrections to Eq.~\eqref{approximated}. In Ref. \cite{gerardone} the
accuracy of this approximation, in the case of inverse power law
waiting time distributions,
was examined, and found to be very good. In this paper we shall make
a discussion of the key results on the aging experiment on BQD
systems, taking into account the error associated to this
approximated formula.

It is possible to predict that the case of
very slow modulation does not yield aging. In the case of a modulated
Poisson process, we replace the model of Eq. (\ref{52})
with
\begin{equation}
\label{modulated}
\frac{dy}{dt} = r(t)
y.
\end{equation}
This means that we set $z= 1$ and we replace the
parameter $\alpha$ with the time dependent rate $r(t)$.
The equation
of motion for $p(y,t)$ is given by
\begin{equation}
\label{poissoncheerfulequation}
\frac{\partial}{\partial t} p(y,t) = r (t)  \left[ -
\frac{\partial}{\partial y} y  p(y,t)  + p(1,t)\right].
\end{equation}
Note that the second term of on the right hand side of this equation
corresponds to the back injection of the particle, when it reaches
the border $y =1$, and thus to the resetting process of the renewal
model of Eq. (\ref{52}).  When $r(t)$ does not depend on time,
Eq.~\eqref{poissoncheerfulequation} represents  a  Poisson process.
Let us focus our attention on the case where $r(t)$ is a stochastic
function of time. If it is very fast, $r(t)$ must be replaced by
$<r(t)>$, the process becomes Poisson again and it departs from the
modulation model adopted in this paper (see Section
\ref{modulation}). Anyway, in accordance to a well known notion, this
Poisson process does not yield aging.  The model of Eq.
(\ref{poissoncheerfulequation}) becomes equivalent to the modulation
model of this paper when the fluctuation of $r(t)$ is very slow.  We
see that the equilibrium distribution coincides with the initial flat
distribution. Thus, we cannot adopt the departure from the initial
distribution as a way to define the system's age. In the case of a
virtually infinitely slow modulation, the system leaves for a
virtually infinite time in a Poisson condition, with no aging
whatsoever.

\section{Aging experiment on artificial
data}\label{artificialdata}

This section is devoted to illustrating,
with the help of artificial data, a technique of analysis aiming at a
quantitative evaluation of the degree of renewal properties of a
given time series. We refer to this kind of analysis as \emph{aging
experiment}.

It is important to notice that the analysis of real
data implies the observation of only one single sequence. In this
case we must turn a single sequence into a very large number of
sequences of the same age. The first sequence is the sequence,
artificial or experimental, to analyze, beginning at time $t = 0$
with the system being located at the beginning of a state, either
``on"  or ``off". The second sequence is obtained from the
first, canceling the first state, namely, shifting the first sequence
towards the time origin by the quantity equal to the time duration of
the first state, so that the second sequence begins at time $t=0$,
when the system begins sojourning in the second state of the first,
or original, sequence. On
the same token, the third sequence begins at time $t=0$, when
the system begins sojourning in the third state, and so on. Thus the
waiting time
distribution $\psi(t)$, $t_{a} = 0$, is the distribution of the time
durations of the first states. To do the aging experiment we set
$t_{a} > 0$ and we record the time lengths of the first states
observed in that time position.
With this prescription we define
$\psi_{t_{a}}(t)$. A quantitative definition of amount of aging is
more properly done using the SP, defined
by
\begin{equation}
\label{definition}
\Psi_{t_{a}}(t) \equiv \int_{t}^{\infty} \psi_{t_{a}}(t^{\prime}) dt^{\prime},
\end{equation}
rather than $\psi_{t_{a}}(t)$.

  From now on we denote by
$\psi^{exp}(t)$ and $\Psi^{exp}(t)$ the waiting time distribution
densities and the SPs, respectively,  derived from the
experimental data. Of course, in the case of artificial data, where
the sequence is realized for the specific purpose of producing the
function $\Psi(t)$ of Eq. (\ref{experimental}) and $\psi(t)$ of Eq.
(\ref{theoretical}), the experimental functions coincide with the
corresponding theoretical prescriptions.
Then, we define the
corresponding aged distributions using
Eq. (\ref{approximated}).
This expression is not exact. However, it is convenient for the
purposes of this paper, where the experimental error is expected to
be larger than the discrepancy between Eq. (\ref{approximated}) and
the exact prescription. Then, we denote with $\Psi^{ren}_{t_{a}}(t)$
the SP derived from the experimental observation,
namely from $\Psi^{exp}_{0}(t) \equiv \Psi^{exp}_{t_{a} = 0}(t) $, by
means of
Eq. (\ref{approximated}).

The prescription adopted in Section
\ref{modulation} to produce the time series $\{\tau_{i}\}$ with a
changing $N_{m}$ for $N_{m} \rightarrow \infty$ becomes coincident
with the slow modulation of Eq. (\ref{poissoncheerfulequation}).
Thus, we expect no aging in this limiting case. In the opposite limit
with $N_{m} = 1$, the
sequence is renewal. Thus, we expect the
maximum amount of aging. In other words, the renewal condition should yield
\begin{equation}
\label{total}
\Psi^{exp}_{t_{a}} (t) = \Psi^{ren}_{t_{a}}(t),
\end{equation}
whereas the condition of very slow modulation should produce no
aging, a property described by:
\begin{equation}
\label{noaging}
\Psi^{exp}_{t_{a}} (t) = \Psi^{ren}_{0}(t) \equiv \Psi_0 ^{exp}.
\end{equation}

   Eqs. (\ref{total}) and (\ref{noaging}) refer to two limiting
conditions: the condition of Eq.(\ref{total}) corresponds to total
aging, with no pseudo events, while the condition of Eq.
(\ref{noaging}) stems from a process dominated by Poisson pseudo
events, with a total lack of aging. It is important to point out that
the  results  presented in Section \ref{realexperiment}, do not
involve  any assumption on the form of the waiting time distribution.

Let us introduce here another important ingredient of our
analysis, the aging intensity function:
%. This important indicator is defined
%by following quantity:
\begin{equation}\label{intensity}
     I_a (\tau)  = \frac{\Psi^{exp}_{t_{a}} (\tau) - \Psi^{exp}_0
(\tau)}{\Psi^{ren}_{t_a}(\tau) -
     \Psi^{exp}_0 (\tau)}.
\end{equation}
Eqs.~\eqref{total} and \eqref{noaging} yield $I_{a}(\tau) = 1$, and
$I_{a}(\tau) = 0$, respectively, thereby indicating that Eq.
(\ref{intensity}) is a proper aging intensity indicator, with $1$ and $0$
representing total aging and lack of aging, respectively. In
principle, this function should decrease from $1$ to $0$ upon increase
of the number of pseudo events.

\begin{figure}
    \subfigure[]{
      \label{pseudo:a}
      \includegraphics[width=8cm]{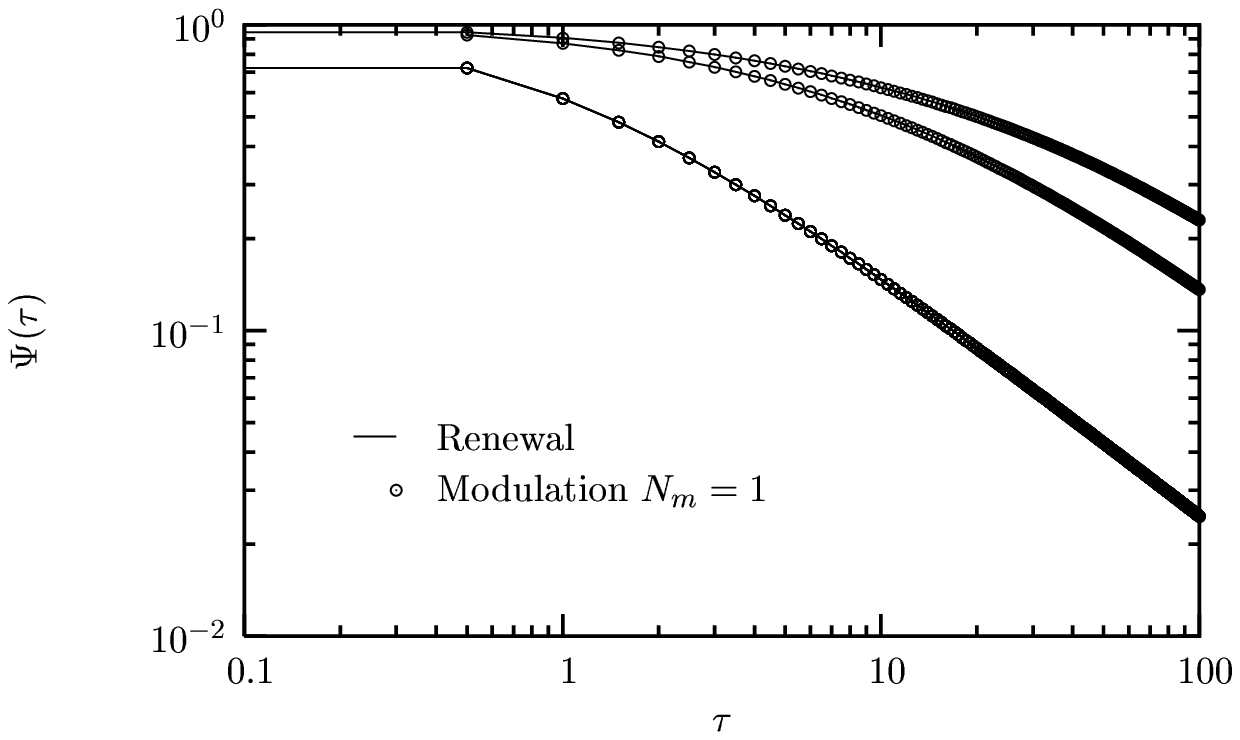}}
    \subfigure[]{
      \label{pseudo:b}
      \includegraphics[width=8cm]{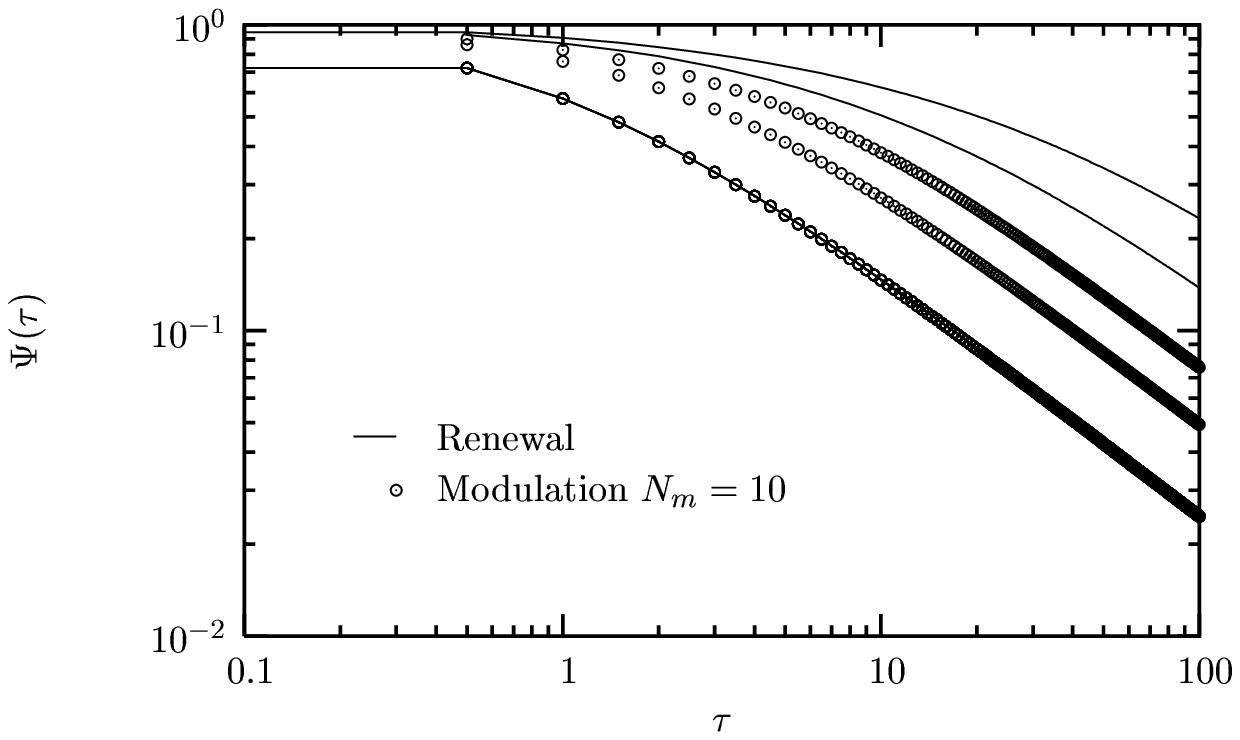}}
    \subfigure[]{
      \label{pseudo:c}
      \includegraphics[width=8cm]{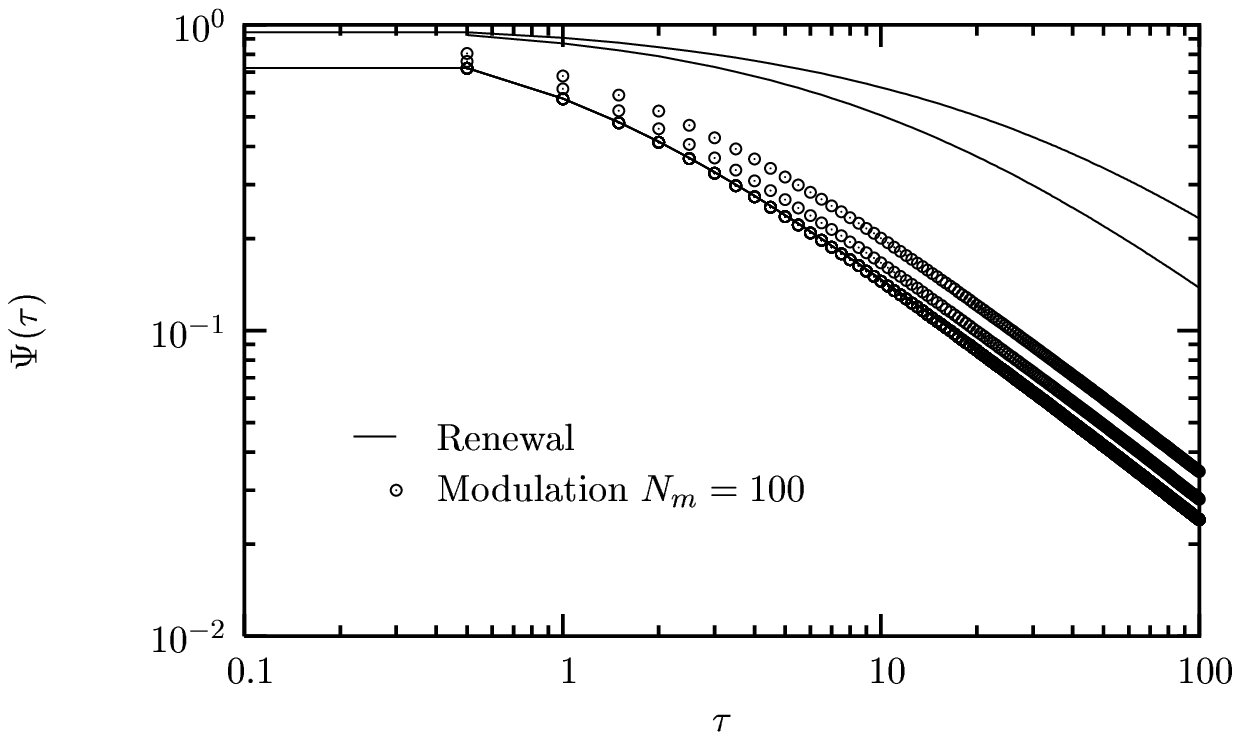}}
\caption{The SP $\Psi_{t_a}^{exp}(t)$ as a function of time. The
three figures
Fig.~\ref{pseudo:a},~\ref{pseudo:b}
and~\ref{pseudo:c} refer
    to $N_{m} = 1$, $N_{m} = 10$ and $N_{m} = 100$, namely to $0$, $9$
and $99$ pseudo events per critical event, respectively. The curves
of each figure are
    obtained from artificial sequences yielding for the waiting time
distribution density the same inverse power law form of
Eq.~\eqref{experimental}, with $\mu = 1.8$. For each figure the three
distinct ages
    $t_a = 0, 20, 60$ (from the bottom to the top) are considered. The
renewal predictions and the modulation results are denoted by
full
lines and circles, respectively. }
    \label{pseudo}
\end{figure}

The numerical results of Figs.~\ref{pseudo:a} to~\ref{pseudo:c}
confirm the expectation that the larger the number of pseudo events,
the smaller the aging intensity. In fact, in accordance with the
earlier theoretical remarks, we see that in the case of no pseudo
event, $N_{m} = 1$, reported in
Fig.~\ref{pseudo:a}, modulation and renewal yield the same amount of
aging. The occurrence of $9$ pseudo events, namely the case $N_{m} =
10$ illustrated in Fig.~\ref{pseudo:b}, is already enough to
significantly reduce modulation aging.
Fig.~\ref{pseudo:c} shows that $N_{m} = 100$ yields an even larger
aging intensity reduction. In conclusion, these
numerical results confirm the theoretical expectation that the
infinitely slow modulation, $N_{m}
= \infty$, should produce no aging.

\begin{figure}[!h]
     \includegraphics[width = 14cm, height =
     6cm]{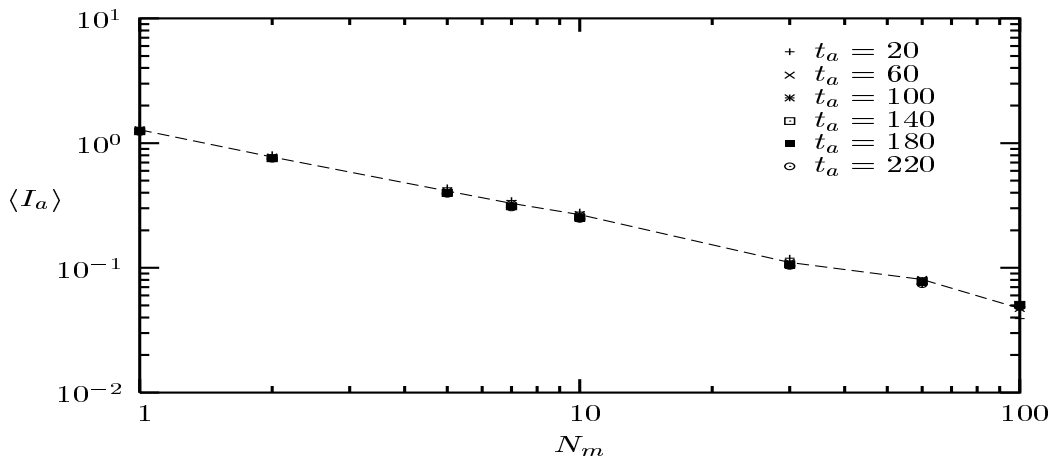}
\caption{The aging intensity indicator $I_{a}(\infty)$ as a function
$N_{m}$, namely $N_{m} -1$ pseudo events per critical event, at
different age, namely different values of $t_a$. The aging intensity
is
defined by Eq.~\eqref{intensity}.  These results refer to artificial
sequences with $\mu = 1.8$. Physical conditions with different ages
and the same number of pseudo events collapse into the same
$I_{a}(\infty)$, thereby becoming indistinguishable in the scale of
this figure. }
     \label{attenuation}
\end{figure}

The adoption of the aging intensity function of Eq. (\ref{intensity}) allows
us to express the aging reduction illustrated by the earlier figures in a
quantitative way. The analysis of the numerical simulations of both renewal
and modulation models shows that the
aging intensity function $I_a(\tau)$ for $\tau \rightarrow \infty$
tends to an asymptotic value,
$I_a(\infty)$. This property is shared by the real data analyzed in
Section \ref{realexperiment}. We estimate this value and we use it,
in both the case of artificial sequences of this section and in the
case of real data of Section \ref{realexperiment}  to define the time
asymptotic intensity of the aging indicator.

Fig. \ref{attenuation} illustrates the application of this procedure
to the case of artificial sequences. We see that the aging intensity
decreases with the increase of the number of pseudo events per
critical event.
   We fit $I_a(\infty)$, as a function of $N_{m}$, with the inverse
power law $(N_m)^{-\alpha}$,
with $\alpha = 0.70 \pm 0.02$.
The aging intensity indicator, $I_{a}(\infty)$, which should hold the
value of $1$ when there are no pseudo events, actually slightly
exceeds this value with no pseudo event and decreases by a factor of
$10$, with increasing the number of pseudo events per critical event
from $0$ to $99$. In Section \ref{realexperiment}, we shall refer to
Fig. \ref{attenuation} to estimate the amount of pseudo events  per
critical event present in the real BQD data.

\begin{figure}[!h]
     \includegraphics[width = 12cm, height =
6cm]{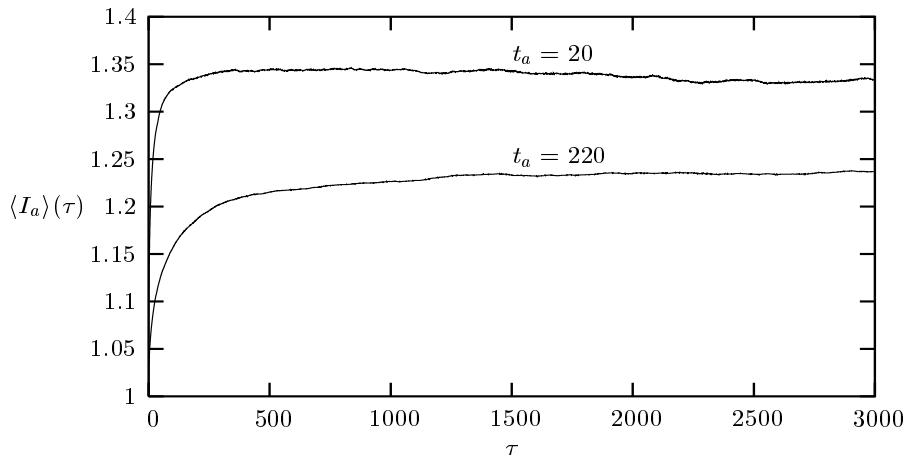}
     \caption{The aging intensity $I_{a}(\tau)$ of
Eq.~\eqref{intensity} as a function of $\tau$. We study the case of
renewal artificial sequences
   with an inverse power law form, with
$\mu = 1.8$. We study the time evolution of this indicator at two
different ages,  $t_a = 20$ and $t_a
     = 220$. The attenuation of the overestimation effect with
     the increasing $t_a$ is evident.}\label{Iatten}
\end{figure}

The aging intensity overestimation, with no pseudo events, is a
consequence of the fact that Eq.~\eqref{approximated} is not exact.
Let us study the effects of this inaccuracy by means of artificial
sequences.   We use artificial sequences derived from the renewal
prescription, namely, by random drawings of numbers from the inverse
power law  distribution
of Eq.~\eqref{theoretical}, yielding a SP of the form of the
Eq.~\eqref{experimental}. We use the values
$\mu = 1.65$ and $\mu = 1.8$, which are typical values of the ``off" 
sequences studied in Section \ref{realexperiment}.  We make a
Monte Carlo simulation and we produce curves of the kind illustrated
in Fig.~\ref{Iatten}, showing the time dependence of the aging
intensity function. We see that $I_a(\tau)$ becomes virtually time
independent after a short
transient.
Let us notice that the levels of these plateaus exceed the
maximum value of $1$.
However, we see that this level becomes closer
to $1$ with the age increase from $t_{a} = 20$ to $t_{a} = 220$. Of
course, the property of a fast attainment of a plateau  is shared
also by the cases, not shown here, of a non vanishing number of
pseudo events, with the plateau height significantly smaller than $1$
for $N_{m}$ of the order of $10$: the results of Fig.
(\ref{attenuation}) have been obtained by evaluating the level of
this plateau. Applying the same criterion to the case of no pseudo
events, we get the asymptotic values $I_a(\infty)$, illustrated by
Table~\ref{errors}.

\begin{table}[!tb]
     \begin{center}
       \begin{tabular}{p{1.2cm}p{1.2cm}p{1.4cm}|p{.3cm} p{1.2cm}p{1.2cm}}
         \hline
         \hline
         & \multicolumn{2}{l|}{\phantom{aaa}$\mu=1.65$} &
\multicolumn{3}{|c}{$\mu=1.8$}\\
         $t_a$ & $I_a(\infty)$  & $\sigma_{ren} (I_a)$ & & $I_a(\infty)$
& $\sigma_{ren} (I_a)$ \\
         \hline
         20 & 1.3734 & 0.0011 & &1.3378 & 0.0047 \\
         60 & 1.31828 & 0.00070 & &1.2765 & 0.0026 \\
         100 &  1.3007 & 0.0012 & &1.2608 & 0.0017 \\
         140 &  1.29325 & 0.00040 & &1.2502 & 0.0014 \\
         180 & 1.28764 & 0.00036 & &1.2420 & 0.0014 \\
         220 & 1.28243 & 0.00037 & &1.2344 & 0.0012 \\
         \hline
         \hline
       \end{tabular}
     \end{center}
     \caption{This table summarizes the results of Monte Carlo
simulations  done to evaluate the aging intensity $I_a(\infty)$.
These results refer to the case of renewal artificial sequences, with
an inverse power law form. The values of $\mu$ adopted, $\mu = 1.65$
and $\mu = 1.8$, are the typical values of the real BQD sequences of
``off" states. $\sigma_{ren}$ is the standard deviation.
%Note that these are asymptotic values (see
%Fig.~\ref{Iatten}).
}
     \label{errors}
\end{table}

In conclusion, the aging indicator $I_a(\infty)$ should be
independent of the system age and equal to $1$: in fact, we are
analyzing artificial sequences  corresponding to the renewal model.
We see, on the contrary, that the aging intensity indicator exceeds
the maximum value of $1$. This is due to the inaccuracy of the
formula of Eq.~\eqref{intensity}, which is known \cite{earlier0}  to
become  accurate for  $t_{a} \rightarrow \infty$. In accordance with
this fact, we see from Table ~\ref{errors} that
  the aging intensity
overestimation tends to vanish with increasing values of $t_{a}$.

\section{Aging experiment on real data}\label{realexperiment}

\begin{figure}[!h]
     \includegraphics[width = 14cm, height = 4cm]{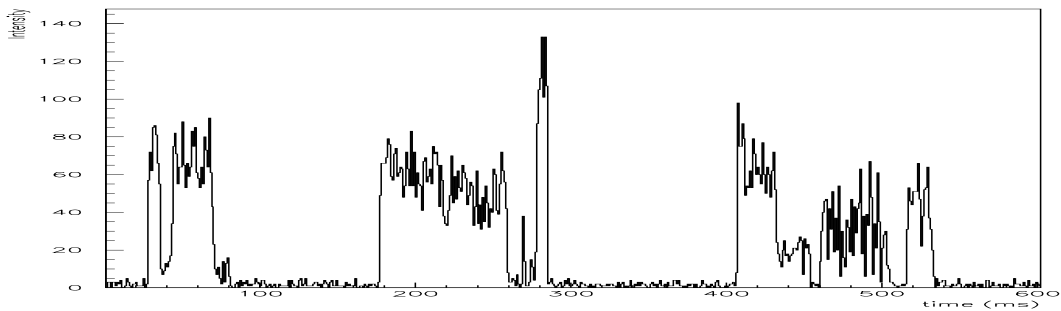}
\caption{A sample of the BQD data examined in this section.}
\label{sample}
\end{figure}

We are now in a position to analyze the experimental results on real BQD data
sequences. The experimental data discussed in this Section have been
obtained by Prof. M. Kuno and V. Protasenko, Dept. of Chemistry and
Biochemistry, University of Notre Dame.
Our data set consists of $32$ sequences of BQD fluorescence
intensities. Each sequence contains $1$ hour of records, sampled every
ms, for a total amount of $360000$ data per sequence. A sample of the
data studied in this section is shown in Fig.~\ref{sample}.

In order to separate the ``on''  from the ``off'' state
it is necessary to define a threshold intensity, above which the
signal corresponds to a `` on'' state and below which it signals
the ``off''  state. We establish this separation along the
lines adopted in an earlier work (see~\cite{nesbitt}), namely with an
iterative procedure for the search of a repartition that make it
possible for  the fluctuations of the two states not to intersect
with the separation line. At the end of this iteration process the
variances of the two states get a well defined value, with $\sigma$
denoting the variance of the ``off'' state: the threshold turns
out to be located at the value
$2\sigma$ over the ``off'' state.

After defining an alternate sequence of ``on" and ``off" states, we 
make the aging experiment with the criterion described
in Section \ref{aging}. Actually, we make three different kinds of
aging experiment. The first and second, considering the waiting times
only of the ``on" and ``off" states, respectively. This
means that we sew the beginning of ``on'' (``off'') state
to the end of the immediately preceding ``on" (``off")
state. The third experiment is done on the whole sequence of waiting
times, with the jump from one state to the other signaling the
presence of an event, whose statistical properties are studied
regardless of whether it corresponds to jumping from the ``on"
to the ``off" state, or vice-versa.

\subsection{Aging of the ``on'' state}
\begin{figure}[!h]
     \includegraphics[angle = 270, scale = 0.5]{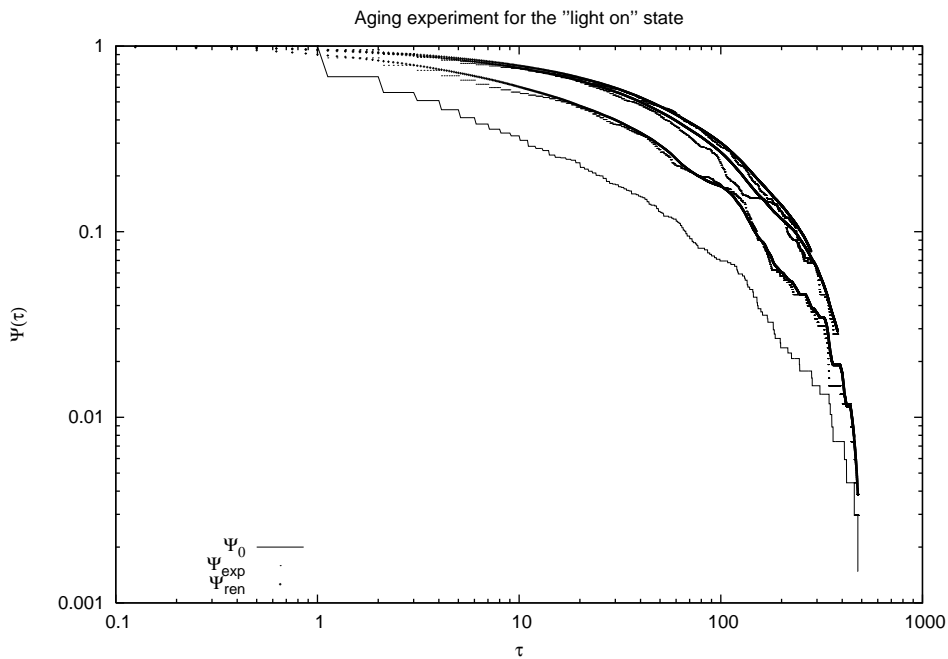}
\caption{The SP $\Psi(t)$ as a function of time.
The experiment is carried using only the ``light on'' waiting time
distribution.  The thin continuous line indicate $\Psi_0^{exp}(t)$,
the dotted line refers to  $\Psi^{exp}(t) $, and the thick continuous
line is $\Psi_{ren}$. From the bottom to the top the curves refer to
$t_a$ with the values $20$, $120$ and $220$.}
\label{agingon}
\end{figure}

Here we discuss the results of the first kind of aging experiments,
   on the
``light on'' state. Fig.~\ref{agingon} shows an example of these aging
experiments. This figure refers to a case where the aging experiment
on the BQD sequence is done for several values of $t_a$, ranging from
$20$ to $220$ in steps of $40$. The thin continuous line represents
$\Psi^{exp}_0$, the dotted line refers to $\Psi^{exp}_{t_{a}}$, while
the thick continuous
line represents $\Psi^{ren}_{t_{a}}$. We see that in this case the
condition of Eq. (\ref{total}) is fulfilled with a very good
accuracy. The systems ages and it does according to the
non-Poisson renewal theory. In fact, aging implies that the system
obeys a non-Poisson statistics, and the fulfillment of  Eq.
(\ref{total}) means that this non-Poisson statistics is renewal.
It is also worth  pointing out again that this observation  does not
involve that the deviation from Poisson statistics is realized
through inverse power laws, as assumed for simplicity in Section
\ref{renewal0}.  In fact, the experimental waiting time distribution
and SP are not inverse power law, or, at least do
not correspond to an inverse power law with a well defined index.
However, they depart from the exponential condition enough as to
generate the aging effects illustrated by Fig.~\ref{agingon}.

\subsection{Aging of the ``off'' state}

\begin{figure}[!h]
     \includegraphics[angle = 270, scale = 0.5]{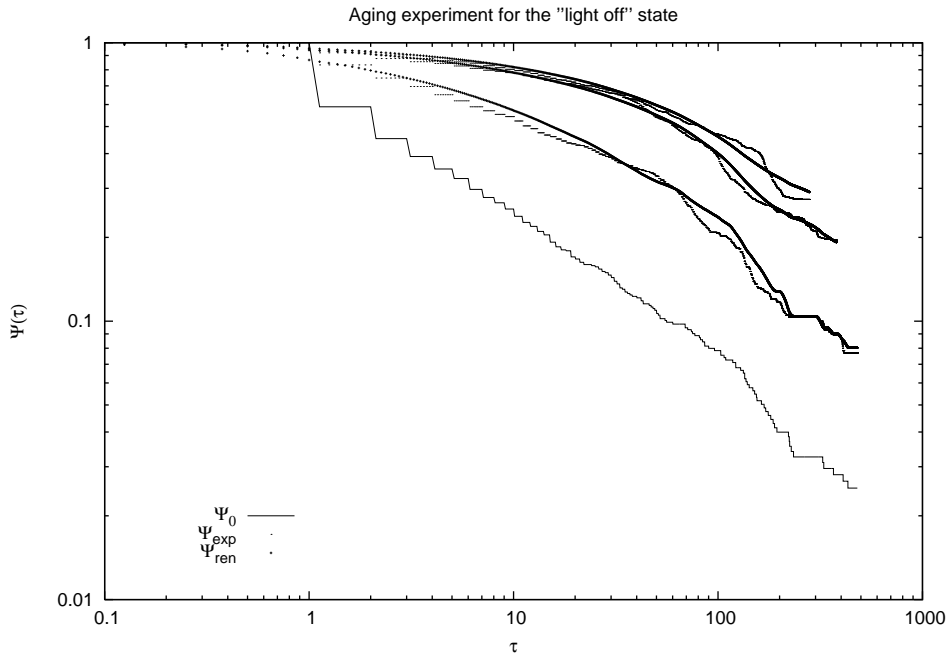}
\caption{The SP $\Psi(t)$ as a function of time.
The experiment is carried using only the ``off'' waiting time
distribution.  The thin continuous line indicate $\Psi_0^{exp}(t)$,
the dotted line refers to  $\Psi^{exp}(t) $, and the thick continuous
line is $\Psi_{ren}$.
    From the bottom to the top the curves refer to $t_a$ with the values
$20$, $120$ and
$220$.}
\label{agingoff}
\end{figure}

Let us discuss  aging for the sequence of waiting times in ``off'' 
state. A sample of these results is shown by
Fig.~\ref{agingoff}. Also
in this case,  the system is aging
and its behavior  is very well described by the renewal theory.

By visual inspection, we can stress some differences between the
result of the latter and the former
experiment. The main difference seems to be that the distribution of
the time of sojourn in ``light on'' state is truncated
after about two decades, while the inverse power law distribution of
the ``off'' states holds
longer. This result confirms the earlier observation of Chung and
Bawendi~\cite{collection}. For this reason, it turns out to be
difficult for us to estimate the index of the waiting time
distribution of the ``on" states. In the case of the ``off"
states,  instead, we can do that,
since the power law behavior is more distinct that in the earlier
case.  Our estimations of the power law exponent, in
this case, ranges from $1.65 \pm 0.02$  to $1.80 \pm 0.05$.
   We note
that the artificial sequences of Table I refer to these values.

\subsection{Aging of the ``on-off'' state}
\begin{figure}[!h]
     \includegraphics[angle = 270, scale = 
0.5]{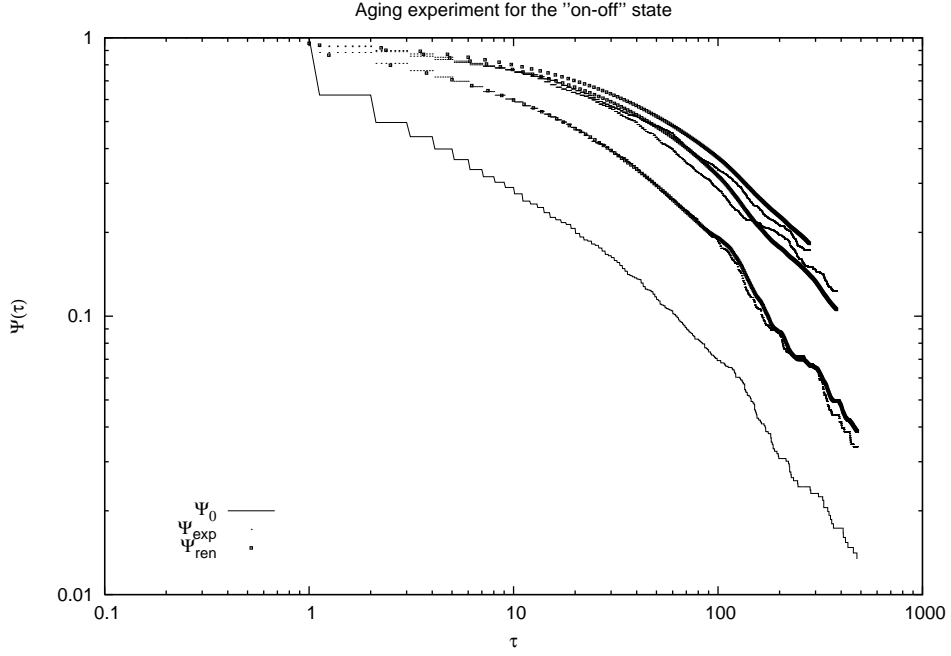}
\caption{The SP $\Psi(t)$ as a function of time.
The experiment is carried using both the ``on'' and ``off"
waiting time  distribution.  The thin continuous line indicate
$\Psi_0^{exp}(t)$, the dotted line refers to  $\Psi^{exp}(t) $, and
the thick continuous line is $\Psi_{ren}$.
    From the bottom to the top the curves refer to $t_a$ with the values
$20$, $120$ and
$220$.}
\label{agingonoff}
\end{figure}

Let us discuss now the result of the third aging experiment, done on
the whole sequence, with the transitions from the ``on" to the
``off" states as the markers of the significant events to
analyze with the aging experiment. Fig.~\ref{agingonoff} shows a
sample of this third kind of aging experiment. Also in this case it
is evident how the theoretical predictions of the renewal theory fit
very well the experimental results. \\ \indent In order to support
quantitatively the conjecture made on the basis of visual inspection
of the results,
in the next subsection we shall adopt the aging intensity indicator
introduced in
Sec.~\ref{artificialdata}.

\subsection{Aging intensity}

We recall that the adoption of Eq. (\ref{approximated}) yields for
the aging indicator $I_a(\infty)$ values larger than $1$, namely, we find
that Eq. (\ref{approximated}) overestimates the aging intensity (see Table
~\ref{errors}). With Fig. 2 we also found that a number of pseudo
events of the order of ten significantly reduces the aging intensity.

\begin{figure}[!h]
     \includegraphics[width = 12cm, height = 6cm]{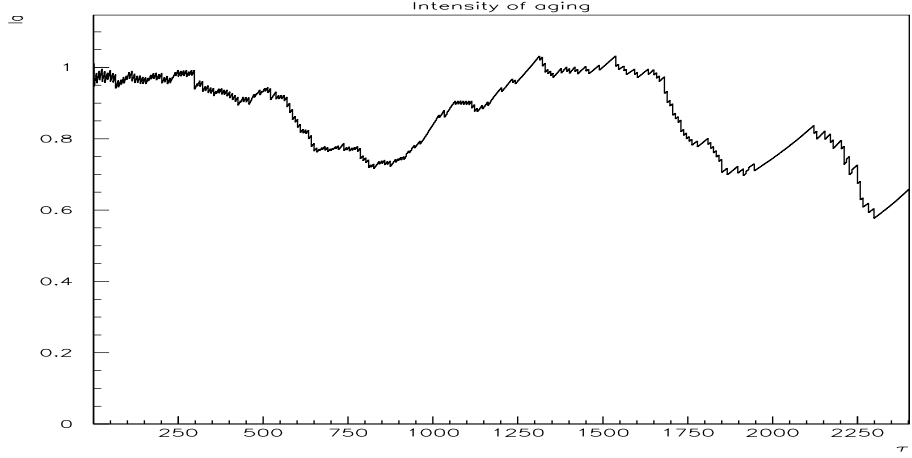}
     \caption{Example of the aging intensity function $I_a(\tau)$ for the
``light on'' state.}\label{esint}
\end{figure}

    We can now use these indications for a rough estimation of the
number of possible pseudo events present in the sequences of sojourn
times in the ``light on" state.  To make the evaluation of the aging
intensity value as statistically accurate as possible, we adopt a
Gibbs ensemble average.
In other words, we make an average over all the sequences at our disposal,
after assigning to them the same age $t_{a}$. This is done by
preparing all the sequences in such a way that at $t= 0$ each of them
begin at the beginning of the time of sojourn in  the ``on"
state,  the ``off" state, or the ``on" or ``off"
state, according to the kind of experiment under study, of the first,
second and third type, respectively. Then we set for all sequences
the beginning of the observation process at the same time $t_{a}$. We
obtain
the mean aging indicator, indicated by $<I_a>$. We do the
same experiment for different values of $t_{a}$. Fig.~\ref{esint}
shows the time evolution of $I_{a}(\tau)$ for a given sequence of "light
on" states corresponding to that used to derive the results of
Fig.~\ref{agingon}. The aging indicator $I_a(\infty)$ for any
sequence of this kind is obtained by making an average on $\tau$, and
the fluctuations around this mean value are used to define the
measurement error.

The average on all the sequences of the Gibbs ensemble are used to
obtain the values of $<I_a>$ reported in the tables.
Tables~\ref{tabagingon},~\ref{tabagingoff} and~\ref{tabagingonoff}
report the value of $<I_a>$ for the ``on'', ``off'' and
``on-off'' waiting time distributions, respectively. In these tables, $\sigma$
represents the maximum likelihood estimation of the standard deviation
of $<I_a>$.

\begin{table}[!tb]
     \begin{center}
       \begin{tabular}{p{1.7cm}p{1.2cm}c}
         \hline
         \hline
        $t_a$ & $<I_a>$ & $\sigma (I_a)$ \\
         \hline
         20 & 0.676 & 0.026\\
         60 & 0.690 & 0.023\\
         100 &  0.810 & 0.021\\
         140 &  0.795 & 0.020\\
         180 & 0.740 & 0.022\\
         220 & 0.868 & 0.019\\
         \hline
         \hline
       \end{tabular}
     \end{center}
     \caption{This table shows the intensity of aging Eq.~\eqref{intensity}
     for the ``light on'' state, for different values of $t_a$, averaged
     over the ensemble. }\label{tabagingon}
\end{table}

\begin{table}[!tb]
     \begin{center}
       \begin{tabular}{p{1.7cm}p{1.2cm}c}
         \hline
         \hline
         $t_a$ & $<I_a>$ & $\sigma (I_a)$ \\
         \hline
         20 & 1.040 & 0.020 \\
         60 & 0.992 & 0.012\\
         100 &  1.030 & 0.014\\
         140 & 1.010 & 0.014\\
         180 & 0.947 & 0.013\\
         220 & 0.962 & 0.015\\
         \hline
         \hline
       \end{tabular}
     \end{center}
     \caption{This table shows the mean aging intensity of Eq.~\eqref{intensity}
     for the ``off'' state, at different values of $t_a$. The
average is carried out on the whole experimental
sequences.}\label{tabagingoff}
\end{table}

\begin{table}[!tb]
     \begin{center}
       \begin{tabular}{p{1.7cm}p{1.2cm}c}
	  \hline
	  \hline
	  $t_a$ & $<I_a>$ & $\sigma (I_a)$ \\
	  \hline
	  20 & 0.949 & 0.013\\
	  60 & 0.894 & 0.011\\
	  100 & 0.914 & 0.011\\
	  140  &0.865  &0.011\\
	  180 & 0.823 & 0.012\\
	  220 & 0.810 & 0.011\\
	  \hline
	  \hline
	\end{tabular}
     \end{center}
     \caption{ This table shows the mean aging intensity
Eq.~\eqref{intensity} for
     the ``on-off'' state, at several values of $t_a$. The average is
carried out on all the experimental sequences.}\label{tabagingonoff}
\end{table}

These results show that the aging of both the ``off'' and ``on-off'' 
waiting time
distributions are very well described by the means of the renewal
theory. The accuracy of the renewal prediction, become worse for the
``light on'' state, especially for  low values of the parameter
$t_a$,  and it improves with the increase of
$t_a$. By comparing these values to the curve of Fig. 2, and taking
into account that with $N_{m} = 1$ the aging intensity indicator
overestimates the aging intensity of the artificial sequences, we
cannot rule out the possibility that the ``light on" state might
involve the presence of from $5$ to $6$ pseudo events. \\

\section{Concluding Remarks} \label{final}
The main result of this paper is the proof that BQD data obey non-Poisson
renewal with a good accuracy. This complexity condition does not stem
from modulation (superstatistics). The statistics of the ``on"
states is not identical to the statistics of ``off" states. The
latter case is closer than the former to the inverse power law
picture of Section \ref{renewal0}.  However, both waiting time
distributions depart significantly  from the exponential form and
produce significant aging effects.

We confirm the general opinion that the
BQD phenomenon obeys renewal prescription, leaving open, however, the
possibility that a finite amount of pseudo events might be involved,
especially for the ``light on" state.
This suggestion emerges from
the results of Tables ~\ref{tabagingon} and ~\ref{tabagingoff} compared to
those of Fig. ~\ref{attenuation}.

We have to
point out that the field of single-system spectroscopy is  wide and
there are examples of process where slow modulation  conditions seem
to apply \cite{klafter1,klafter2,YanCao02}. This paper affords
prescriptions of statistical analysis that might turn out to be
useful to study the unknown territory between totally renewal and
infinitely slow modulation.

\acknowledgments{We thankfully
acknowledge Prof. M. Kuno and V. Protasenko for kindly making their
experimental data available to us.
S. B. and P. G. thankfully acknowledge financial support from Welch
through grant \# 70525. P. P. thankfully acknowledges financial
support from MIUR through grant ``bando $1105/2002$'', project $\#
245$ and CNR 2005 short term mobility program. }

\end{document}